\documentclass[conference]{IEEEtran}
\IEEEoverridecommandlockouts
\usepackage{cite}
\usepackage{amsmath,amssymb,amsfonts}
\usepackage{amssymb}
\usepackage{mathrsfs }
\usepackage{algorithm}
\usepackage[noend]{algpseudocode}
\newtheorem{mydef}{Definition}
\newtheorem{myth}{Theorem}

\makeatletter
\def\BState{\State\hskip-\ALG@thistlm}
\makeatother
\usepackage{graphicx}
\usepackage{subfig}
\usepackage{textcomp}
\usepackage[many]{tcolorbox}
\usepackage{xcolor}
\usepackage[font=small]{caption}
\def\BibTeX{{\rm B\kern-.05em{\sc i\kern-.025em b}\kern-.08em
    T\kern-.1667em\lower.7ex\hbox{E}\kern-.125emX}}
\begin{document}

\title{Energy Concealment based Compressive Sensing Encryption  for Perfect Secrecy for IoT \\
}

\author{\IEEEauthorblockN{Gajraj Kuldeep, Qi Zhang}
\IEEEauthorblockA{{DIGIT, Department of Engineering, Aarhus University, Denmark}\\
Email:\{gkuldeep, qz\}@eng.au.dk}\\
\textit{Accepted in IEEE Global Communications Conference 2020}
}

\maketitle

\begin{abstract}
Recent study has shown that compressive sensing (CS) based computationally secure scheme using Gaussian or Binomial sensing matrix in resource-constrained IoT devices is vulnerable to ciphertext-only attack. Although the CS-based perfectly secure scheme has no such vulnerabilities, the practical realization of the perfectly secure scheme is challenging, because it requires an additional secure channel to transmit the measurement norm.
In this paper, we devise a practical realization of a perfectly secure scheme by concealing energy in which the requirement of an additional secure channel is removed. Since the generation of Gaussian sensing matrices is not feasible in resource-constrained IoT devices,  approximate Gaussian sensing matrices are generated using linear feedback shift registers. We also demonstrate the implementation feasibility of the proposed perfectly secure scheme in practice without additional complexity. Furthermore, the security analysis of the proposed scheme is performed and compared with the state-of-the-art compressive sensing based energy obfuscation scheme.
  
\end{abstract}


\section{Introduction}

The Internet of things (IoT) has expedited the process of doing many vertical tasks by interconnecting diverse technologies such as wireless sensor networks, embedded systems, control systems, automation, and other technologies. This evolution has resulted in the exponential growth of data generation and a multifold increase in IoT devices.  The ever increasing  data and resource-constrained nature of IoT devices have posed critical challenges such as efficient data representation, data transmission and information secrecy. 

Compressive sensing (CS) provides data compression by reducing the number of samples using a sensing matrix \cite{bib:EJT,bib:DLD,bib:EJMB}.   With the proper selection of a sensing matrix, CS can also provide obfuscation \cite{bib:EJTT}. The computational secrecy of the CS measurements for Gaussian sensing matrices has been studied in \cite{Rachlin} for one-time sensing. However, computational secrecy requires large-signal length that imposes challenges to achieve computational secrecy in resource-constrained IoT devices~\cite{OurP}. Authors\cite{prefect} prove that one-time sensing of constant energy signals is perfectly secure if the elements of the sensing matrix is Gaussian distributed. The indistinguishability of CS measurements has been studied in  \cite{prefect, Ind} for Gaussian and Binomial sensing matrix.

 CS-based perfectly secure scheme is a promising candidate for simultaneous data compression and encryption\cite{cse1}. However, practical implementation of perfectly secure scheme is challenging because of the requirements of constant energy signal and Gaussian sensing matrix for each measurement. For a general class of signals, a perfectly secure scheme is realized by normalizing the measurements and sending measurement norm through a secure channel \cite{prefect}.  In this way, this scheme requires an additional secure channel, which necessitates a cryptographic algorithm, key management for this algorithm, and transmission of the measurement norm. Therefore,  these issues make this scheme  infeasible to be used in most of the resource-constrained IoT devices. 
  
Our contributions in this paper are as follows. (i) We propose a novel energy concealment encryption scheme by introducing a variable for energy concealment. The proposed scheme tackles the challenges mentioned above. (ii) A cryptographic primitive is designed by utilizing the combination of linear and non-linear feedback shift registers to construct sensing matrices. (iii) We compare the performance of our scheme with the one-time sensing scheme. Although the plaintext attack on the Binomial sensing matrix is conventionally regarded as computationally infeasible~\cite{p1}, we prove that it is prone to cryptanalysis using a pair of plaintext consisting of super increasing sequence and its corresponding ciphertext. (iv) We also demonstrate that our proposed scheme is resistant against the chosen plaintext and ciphertext attack. We further show that the energy obfuscation scheme~\cite{EOS} is also vulnerable to ciphertext-only attacks. (v) We prove that retrieving the key from our designed sensing matrix generating sequence is equivalent to signal separation problem and is more cumbersome than the brute-force attack.

The paper is organized as follows: Section II describes the theoretical aspect of the CS-based encryption schemes. Section III explains the proposed energy concealment encryption scheme and its performance evaluation. Section IV presents security analysis of the proposed scheme. Finally Section V concludes the paper. 

\textit{Notations:} In this paper, all the boldface uppercase and all the boldface lowercase  letters represent matrices and vectors, respectively. $\mathbf{x}^T$ is transpose of $\mathbf{x}$.
The italic letters represent variables. 
\section{Theoretical security of CS-based schemes }

Let plaintext $\mathbf{x}\in \mathbb{R}^N$ be an $N$-dimensional column vector. A vector,  $\mathbf{x}$, is called $K$-sparse if there exists a transform,  $\mathbf{\Psi}$, such that:
\begin{eqnarray}
\mathbf{x=\Psi \theta}
, \label{sparseT}
\end{eqnarray}
and $\mathbf{||\theta||_0}\le K$. A signal is called compressible if it has representation as shown in Eq. \ref{sparseT} with a few large coefficients and many small coefficients in the transformed domain. A compressible signal can be approximated to $K$-sparse signal, which makes the CS encoding and decoding similar for compressible and $K$-sparse signals  \cite{bib:EJT,bib:DLD,bib:EJMB}. By applying the CS encoding on the $i^{th}$ block of plaintext, $\mathbf{x}_i$, we get the $i^{th}$ measurement vector, $\mathbf{y}_i$, as,
\begin{eqnarray}
\mathbf{y}_i=\mathbf{\Phi}_i \mathbf{x}_i
,\label{CSeqn}
\end{eqnarray}
where $\mathbf{y}_i\in \mathbb{R}^M$ is called ciphertext and  $\mathbf{\Phi}_i \in \mathbb{R}^{M\times N}$ is a sensing matrix.  Encryption scheme based on Eq. \ref{CSeqn} is called one-time sensing (OTS) if $\mathbf{\Phi}_i$ is changed for every new measurement.

The reconstruction of the plaintext is performed by the decoding algorithm as given below,
\begin{eqnarray}
\mathbf{\hat{x}}_i=\arg \min \limits_{\mathbf{x}_i \in \mathbb{R}^N} \mathbf{||x}_i||_1 \text{       s.t.   } \mathbf{y}_i=\mathbf{\Phi}_i \mathbf{\Psi} \mathbf{\theta}_i, \label{l1rec} 
\end{eqnarray} 
where  $\hat{\mathbf{x}}_i$ is the reconstructed plaintext. Eq. \ref{CSeqn} and \ref{l1rec} are CS-based encryption and decryption, respectively. Note that if signal is $K$-sparse in the canonical form then the transform, $\mathbf{\Psi}$, is identity matrix in Eq. \ref{l1rec}.

  CS encryption scheme realized using Eq. \ref{CSeqn} is  computationally secure under the condition of OTS and Gaussian distributed sensing matrices  \cite{Rachlin}. Perfect secrecy of CS-based encryption  is studied in \cite{prefect,Ind,Ind1}. Mutual information \cite{prefect} between  a pair of plaintext and ciphertext when using OTS and Gaussian distributed sensing matrix is given as,
  \begin{eqnarray}
 I(\mathbf{x}_i;\mathbf{y}_i)&=&I(E_{\mathbf{x}_i};\mathbf{y}_i),\nonumber \\
 &=&I(E_{\mathbf{x}_i};E_{\mathbf{y}_i}), \label{energy}
 \end{eqnarray} 
 where $E_{\mathbf{x}_i}$ and $E_{\mathbf{y}_i}$ are energy of $\mathbf{x}_i$ and $\mathbf{y}_i$, respectively. 
 
  Eq. \ref{energy}  guarantees asymptotic perfect secrecy for constant energy input signals. As we know that most of the practical signals do not have constant energy. The CS-based encryption scheme
  suggested in \cite{prefect} is perfectly secure for general signals. The authors used the following strategy at the transmitter to achieve perfect secrecy, 
 \begin{equation} \label{norma}
 \mathbf{y}'_i =
 \begin{cases}
 \frac{\mathbf{y}_i}{||\mathbf{y}_i||_2} \hspace{1.35cm} \text{       if }||\mathbf{y}_i||^2_2 >0,\\
 \mathbf{u} \hspace{2cm}     \text{         else},   
 
 \end{cases} 
 \end{equation}
 where $\mathbf{y}_i$ is the $i^{th}$ block of ciphertext, $\mathbf{y}'_i$ is the $i^{th}$ block of normalized ciphertext and $\mathbf{u}$ is a random vector with uniform distribution and unit norm. $ \mathbf{y}'_i$ is transmitted through insecure channel and its corresponding norm $||\mathbf{y}_i||_2$ is transmitted through an additional secure channel. Therefore, a separate encryption algorithm is required for the transmission of the norm. We can observe that there are some challenges in the practicality of the suggested scheme. 
 (i) For a signal of all zeros, the measurement vector is also a zero vector, which leads to the detection of the actual zero signal by the adversary. Therefore, in the perfectly secure scheme a uniform distributed random vector is sent instead of the measurements of zeros, which in turn makes the detection of the zero signal at receiver impossible. 
 (ii) An addition secure channel is required for the transmission of the measurement norm. It requires an additional encryption algorithm and key management at the IoT device. Furthermore, transmission of the measurement norm causes extra burden on the communication protocol.
 (iii) The existing perfectly secure scheme cannot exploit robustness provided by the compressive sensing. Namely, error in the estimated energy at the receiver can result in deviation in the reconstructed signal. In the next section, we propose the energy concealment encryption scheme which tackles the above-mentioned challenges in the existing perfectly secure scheme.  

\section{Energy concealment based CS encryption scheme}
Energy concealment encryption scheme is realized  by inserting an energy concealing variable in the input data vector and  designing a novel cryptographic primitive for the construction of Gaussian sensing matrices. 
\subsection {Constant energy signal construction}

Let the input data vector $\mathbf{x}$ be of length $N-1$. Assuming the maximum energy of input vector  is $E_{max}$. We construct a new vector $\mathbf{x}'$ by adding an energy concealing variable $c$ to $\mathbf{x}$ such that energy of $\mathbf{x}'$ is $E_{max}$.
 $\mathbf{x}$ is given as:
\begin{eqnarray}
\mathbf{x}=[x_1,x_2,\dots,x_{N-1}]^T. \label{vect}
\end{eqnarray}
The energy concealing variable, $c$, is created as:
\begin{eqnarray}
c=\sqrt{E_{max}-||\mathbf{x}||_2^2} \label{const}
\end{eqnarray}

Signal $\mathbf{x}'$ is constructed by concatenating  $c$ with $\mathbf{x}$ as given in Eq. \ref{vect}. We get 
\begin{eqnarray}
\mathbf{x}'=[c,x_1,x_2,\dots,x_{N-1}]^T \label{nvect}.
\end{eqnarray}

From Eq.  \ref{const} and \ref{nvect}, it is clear that the new signal, $\mathbf{x}'$, has constant energy, i.e. $E_{max}$.  From Eq. \ref{energy} we know that the measurements leak only the signal energy. If the maximum energy of the signal is made public, the adversary does not learn anything new from the measurements. 

It is clear that the proposed scheme does not require an addition secure channel because $\mathbf{x}'$ can be now encrypted using Eq. \ref{CSeqn}.  
Transmission of a zero vector is tackled by sending the non-zero vector. It becomes difficult for adversary to distinguish a zero input vector from a non-zero input vector, because the sensing matrix is changed for each input. 
Moreover, the proposed scheme maintains the inherent robustness provided by CS, because there is no post-processing on measurement data. In a nutshell, this scheme provides a solution to simultaneously address all the challenges described in Section II. Note that all the described results of perfect secrecy are valid only if the sensing matrix is Gaussian distributed and not used more than once. Therefore, a remaining challenge is the construction of Gaussian sensing matrices.

Since generating a truly Gaussian distributed sensing matrix is not possible in practice, we propose a method to generate approximately Gaussian distributed sensing matrices. Energy concealment encryption scheme can be compared with the symmetric key algorithms for the case when the sensing matrix is changed for each measurement, and sensing matrices are constructed using a pseudorandom number generator. 

\subsection{Signal decoding}
\begin{figure}[htbp]
	\centering
	\subfloat[]{	\includegraphics [width=.5\linewidth]{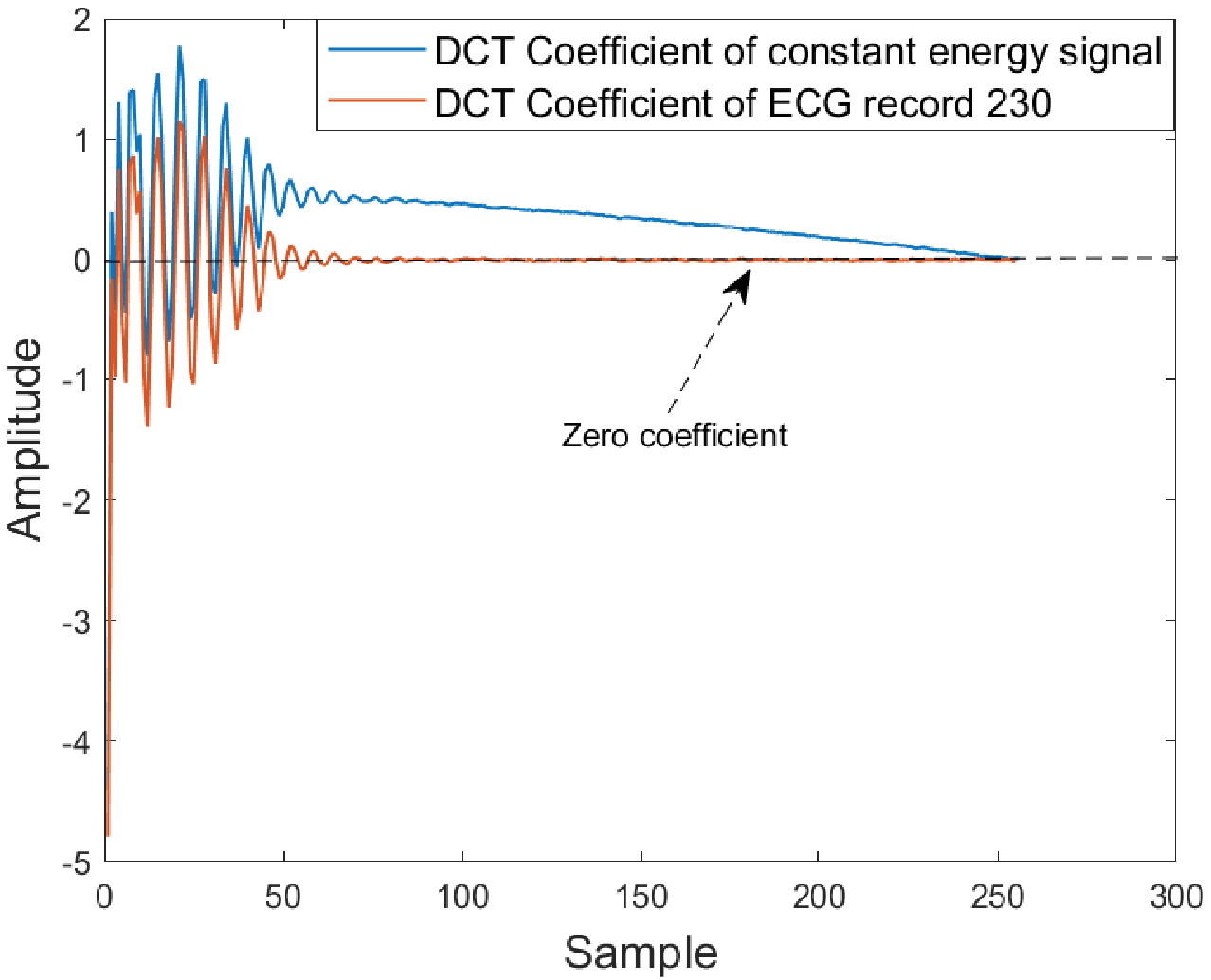}
		
		\label{SC}}
	\subfloat[]{\includegraphics [width=.4\linewidth]{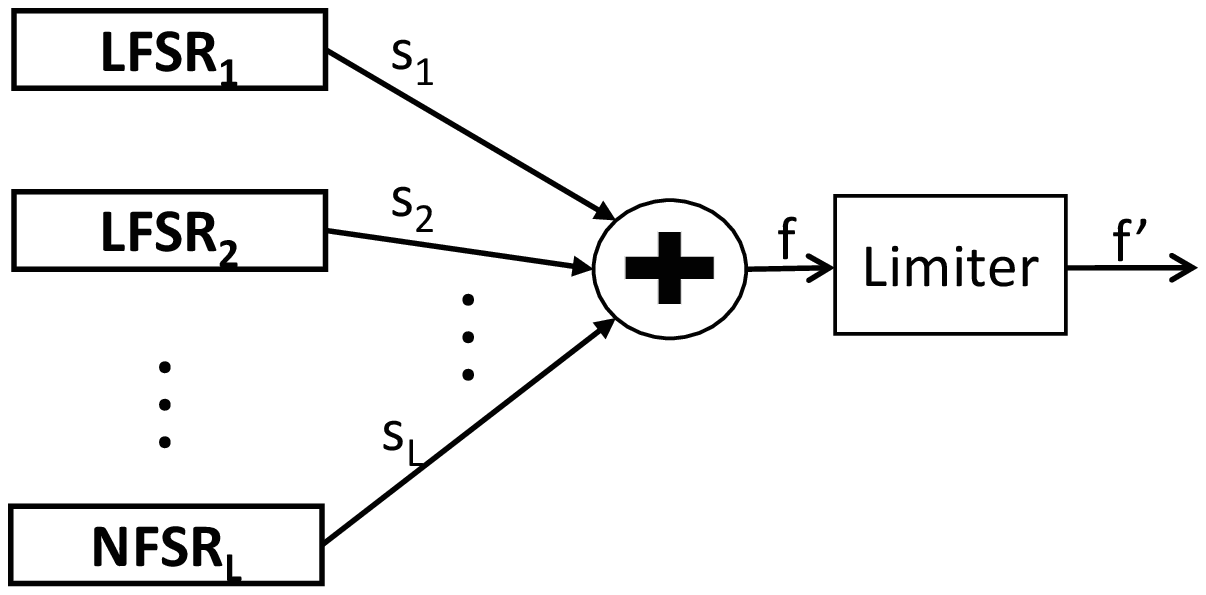}
		
		\label{Gauss2}}
	\caption{(a) DCT transformed ECG signals. (b) Sensing matrix generation framework.}
	
\end{figure}

The insertion of an energy concealing variable to the input signal may change the sparsity basis. Let's assume that the input signal is sparse in $\mathbf{\Psi}$ basis. But, it may not remain sparse in $\mathbf{\Psi}$  after the insertion of the energy concealing variable.  For example, ECG signal is sparse in the discrete cosine transform (DCT) domain and to make it a constant energy signal, a variable is introduced as described in Eq. \ref{nvect}. For constant energy signals, change of sparsity can be seen in Fig. \ref{SC}. It can be observed that the  ECG signal with constant energy is no longer sparse in the DCT domain. This problem has been already tackled by decoding using two orthogonal bases without increasing the number of measurements \cite{DataG, DataG1}. The signal with constant energy can be represented as:
		\begin{eqnarray}
		\mathbf{x}'=\mathbf{x}_{\mathbf{B}_1}+\mathbf{x}_{\mathbf{B}_2}, \label{twoo}
		\end{eqnarray}
		where $\mathbf{x}_{\mathbf{B}_1}$ is sparse in $\mathbf{B}_{1}$ basis and $\mathbf{x}_{\mathbf{B}_2}$ is sparse in $\mathbf{B}_2$ basis.
		From Eq. \ref{twoo} we obtain,
	
	$	\mathbf{x}_{\mathbf{B}_1}=[0,x_1,x_2,\dots,x_{N-1}]^T 
		\text{ and }
		\mathbf{x}_{\mathbf{B}_2}=[c,0,0,\dots,0]^T$. 
Therefore, the decoding for energy concealment encryption scheme is performed by the combined basis of the DCT transform matrix and identity matrix.		
	
\subsection{Design of sensing matrix}
 In this subsection, we show that the Gaussian sensing matrix can be generated using the combination of linear feedback shift registers (LFSR) and a nonlinear feedback shift register (NFSR) as shown in Fig. \ref{Gauss2}.
 From the probability theory, we know that the Binomial distribution can be approximated to the Gaussian distribution. An LFSR generates pseudo random sequence, and its bits can be regarded as Bernoulli distributed with probability $\frac{1}{2}$ \cite{prob}. The Binomial distribution is the sum of Bernoulli distributed random variables as given below,
\begin{eqnarray}
f_i = s_{1_i}+s_{2_i}+\dots+s_{L_i}, \label{Gauss1}
\end{eqnarray}
   where $f_i$ is Binomial distributed and $s_{j_i}$s are Bernoulli distributed with probability $p=\frac{1}{2}$. 
    For $L\ge 10$ and $p=\frac{1}{2}$, $f_i$ can be approximated to Gaussian distribution with mean $Lp$ and variance $Lp(1-p)$. For designing Gaussian sensing matrix using the bits generated by the method described in Fig. \ref{Gauss2}, each LFSR and NFSR output $0$ is converted to $-1$. Therefore, $f_i$ in Eq. \ref{Gauss1} is approximately Gaussian distributed with mean $0$ and variance $L$. To have Gaussian distribution with variance one, Eq. \ref{Gauss1} should be normalized with $\sqrt{L}$. 
    The sequence generated using Eq. \ref{Gauss1} follows Gaussian distribution and has its values from the set generated using $L-2i$ for $i=0$ to $L$. For $L=11$ the possible values are $\{-11,-9,-7,-5,-3,-1,1,3,5,7,9,11\}$. The extreme values such as $11$ and $-11$ reveal the LFSR sequences. The limiter function is used to remove the extreme values. The limiter function is defined as:
 
    \begin{equation} \label{Gauss3}
    f_i'=Limiter(f_i)= 
    \begin{cases}
    f_i \hspace{2cm} \text{if } |f_i|\ne L \\
    Null \hspace{1.5cm}\text{else}.  
    
    \end{cases} 
    \end{equation}
    
    The $Null$ output of the limiter does not mean zero value. It means no value is passed for further processing, in short, skipping one clock cycle of the circuit for each $Null$ occurrence. 
    In this case the sensing matrix will contain values from the set generated by  $L-2i$ for $i=1$ to $L-1$. Gaussian sensing matrix is constructed by taking $MN$ samples of Eq. \ref{Gauss3} and arranging into an $M\times N$ matrix. We represent sequences generated using Eq. \ref{Gauss1} and \ref{Gauss3} as   $\mathscr{F}$ and $\mathscr{F'}$, respectively. One can notice that for $L=3$ sequence $\mathscr{F'}$ is binary. Therefore, a Binomial sensing matrix can also be constructed.  In the next section, the performance of the proposed scheme is evaluated by considering the introduction of an energy concealing variable and the approximately Gaussian distributed sensing matrix. 
    


\subsection{{Signal recovery evaluation}}
\begin{figure}[htbp]
	\centering
	\subfloat[]{	\includegraphics [width=.48\linewidth]{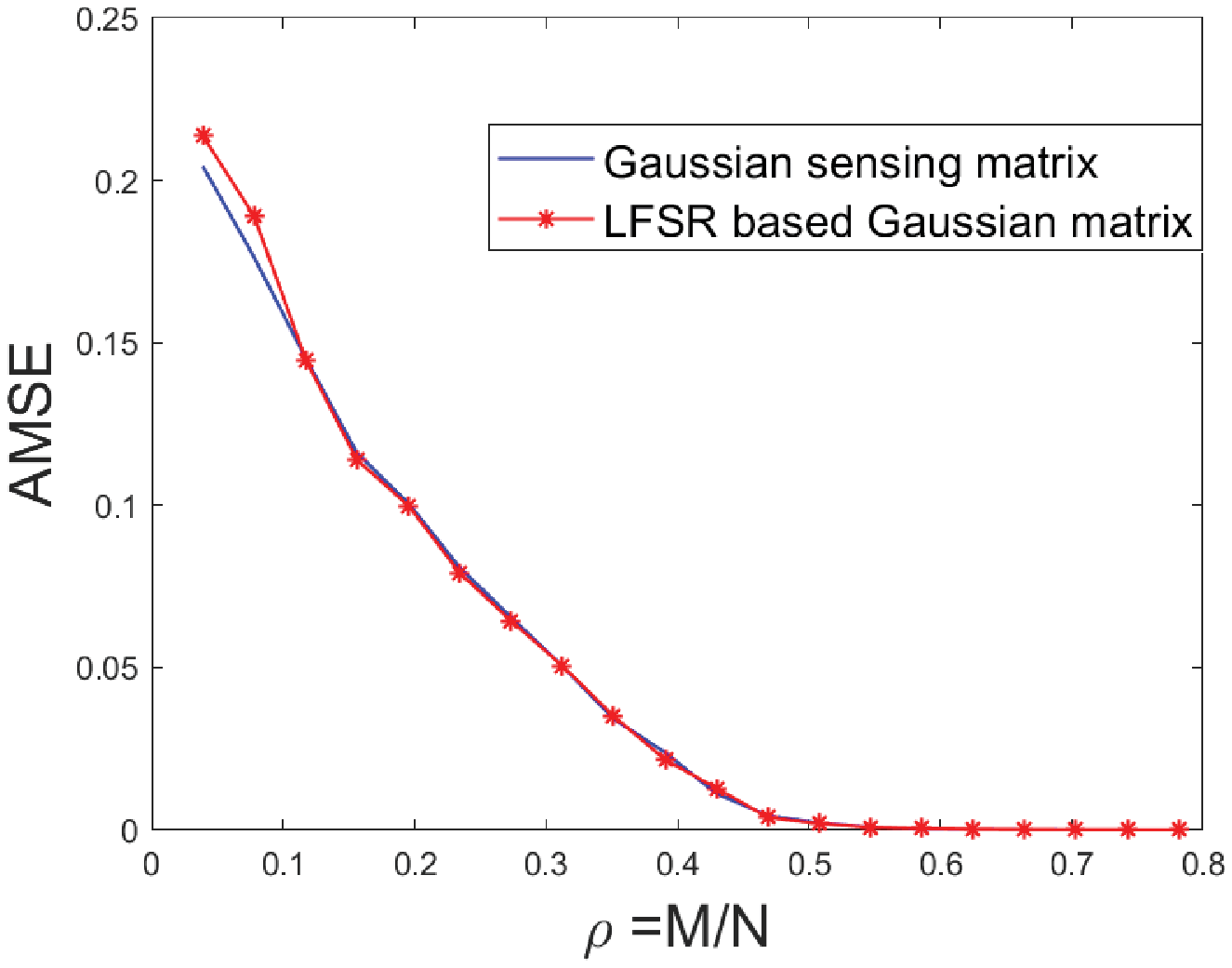}
		\label{mse}}
	\subfloat[]{	\includegraphics [width=.48\linewidth]{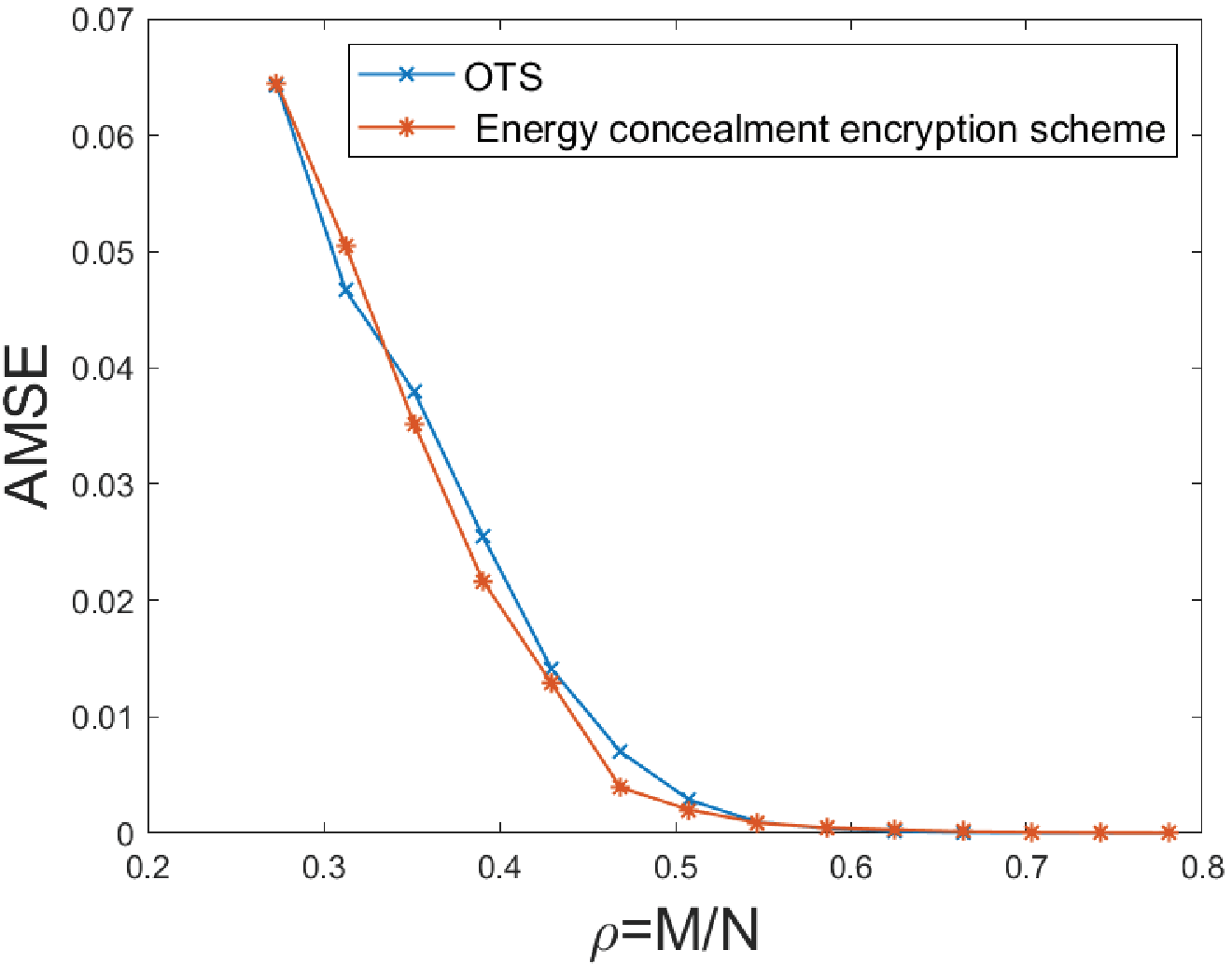}
		\label{recovery}}
	\caption{(a) Error performance for Gaussian and LFSR based sensing matrix. (b) Average error performance for OTS scheme and energy concealment encryption scheme for 100 plaintext blocks.}
\end{figure}

Sensing matrix follows approximately Gaussian distribution; hence, the bound on the number of measurements is same as for the Gaussian sensing matrix. We demonstrate the performance of the proposed scheme using electrocardiogram (ECG) signal samples  from MIT-BIH Arrhythmia Database~\cite{database}. The input length, $N$, is fixed to 256. Error performance of the Gaussian sensing matrix and the sensing matrix generated using LFSRs described in Section III-B is compared for ECG record 230 for various sparsity values, as shown in Fig. \ref{mse}. The average mean square error (AMSE) is calculated between the original and reconstructed signal for 50 iterations. From Fig. \ref{mse}, it is clear that using shift register based sensing matrix does not cause any degradation in the reconstructed signal.

 For our setting, $\mathbf{x}_{\mathbf{B}_1}$ is ECG signal, and it is sparse in DCT basis. Whereas $\mathbf{x}_{\mathbf{B}_2}$ is sparse in Identity basis. AMSE is compared for the OTS with reconstruction using DCT basis and energy concealment encryption scheme with reconstruction using two orthonormal bases for 100 plaintext blocks of each size 256. The reconstruction performance is shown in Fig. \ref{recovery}. From the figure, it is clear that the reconstruction using two orthonormal bases has similar performance as the conventional reconstruction using DCT basis for $\rho \ge 0.3$,  which is determined by the number of non-zero DCT coefficients of the signal.

\section{Security analysis}
In this section, attacks on the energy concealment encryption scheme  and state-of-the-art schemes are discussed. 

\subsection{Ciphertext-only attack}
\begin{figure}
 	\centering
 	\subfloat[]{\includegraphics [width=.15\textwidth] {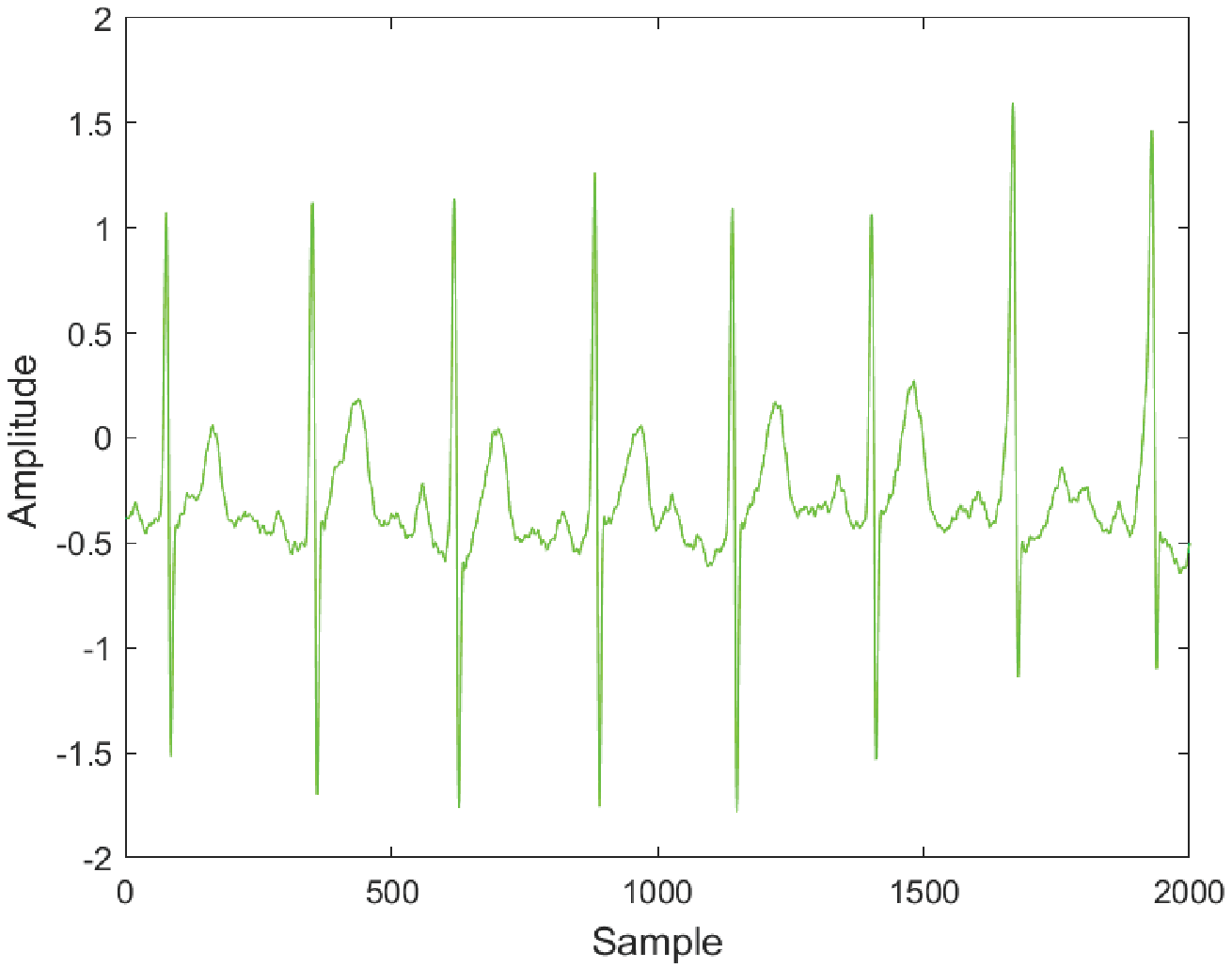}	\label{ca1}}
 	\subfloat[]{\includegraphics [width=.15\textwidth]{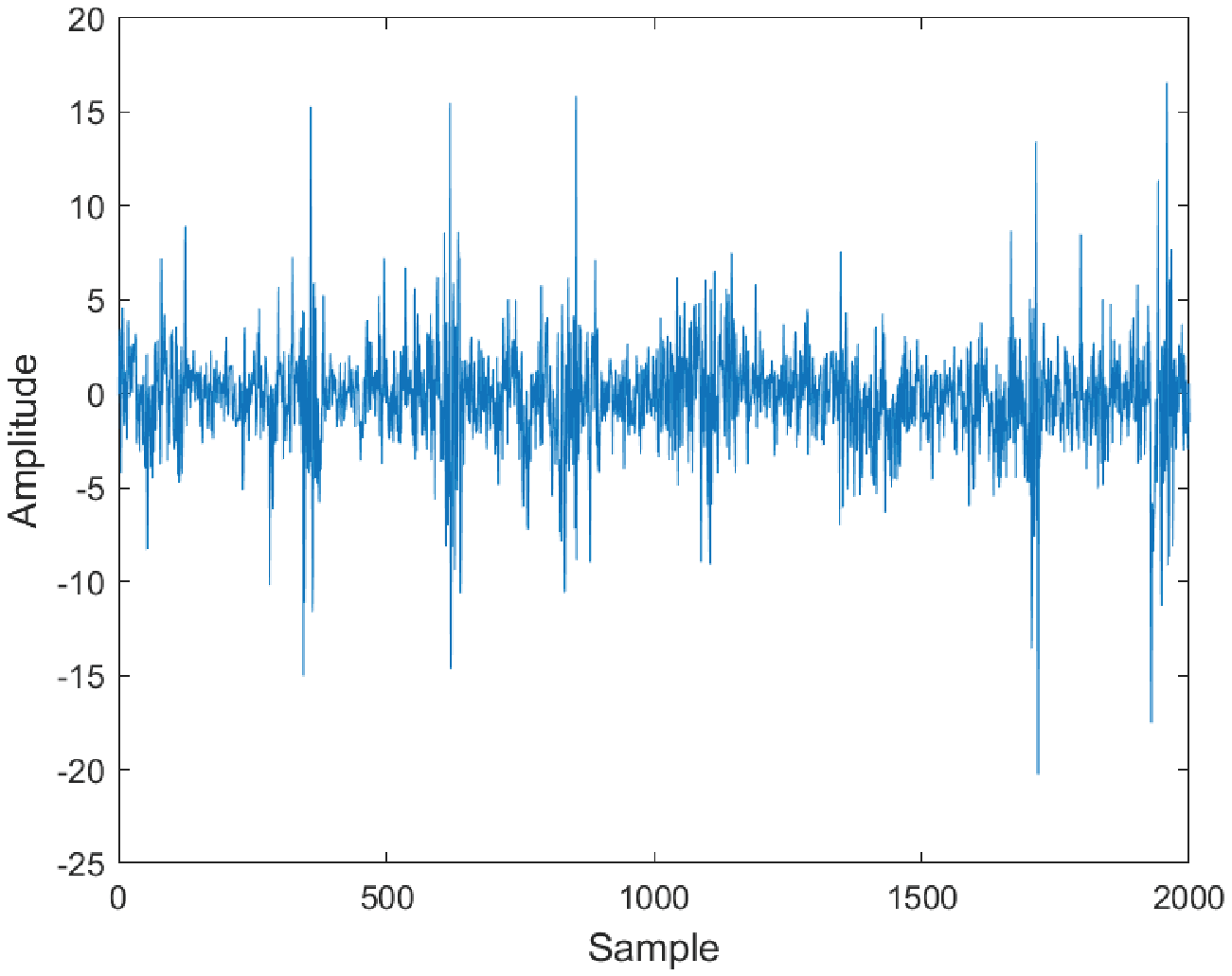}\label{ca2}}
 	\subfloat[]{	\includegraphics [width=.15\textwidth]{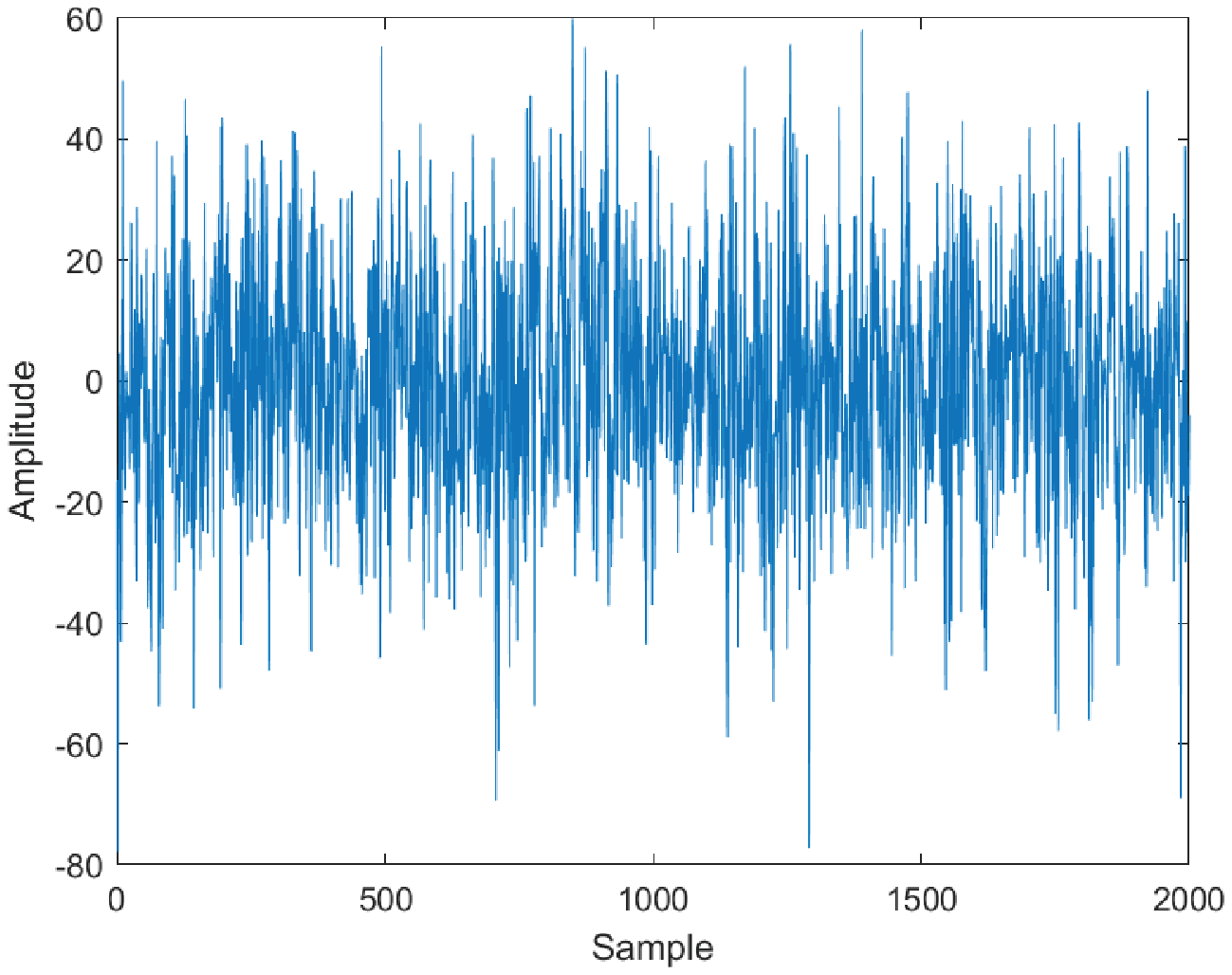}\label{ca3}}
 	\caption{Application of ciphertext-only attack on computationally secure and energy concealment encryption scheme. (a) Original ECG signal. (b) Reconstructed signal from OTS scheme~\cite{OurP}. (c) Reconstructed signal from energy concealment encryption scheme.}
 \end{figure}

Ciphertext-only attack is applied with the following assumptions:
\begin{itemize}
	\item The period of the pseudo-random generator is sufficiently large. The keys of the pseudo-random generator are changed before the repetition of sequences. In this case, the ciphertext-only attack is limited to the analysis of the measurements.
	\item The maximum signal energy is known to the public.
\end{itemize}
   From Eq. \ref{energy}, it is clear that only the energy of the signal is leaked to the adversary. In the energy concealment encryption scheme, the energy is made constant by embedding a variable to the input signal. Therefore, the measurements does not leak any information about the signal. The adversary will acquire an $M$-sparse plaintext instead of a $K$-sparse plaintext if reconstruction is performed using different sensing matrix. Reconstruction using different sensing matrix results in large reconstruction error, and it increases as the number of measurements are increased \cite{Rachlin}.  Authors \cite{Ademors} analyzed the structural attacks on the CS-based encryption schemes,  and concluded that  it is computationally infeasible for an adversary to apply them. Recently, it is shown that the ciphertext-only attack is possible to the computationally secure scheme for small signal lengths \cite{OurP}. However, this attack is not applicable to perfectly secure scheme. Application of this ciphertext-only attack on the computationally secure scheme is shown in Fig. \ref{ca2} for $N=32$ on ECG record 230. From the figure, it is clear that the reconstructed signal follows the original signal, as shown in  Fig. \ref{ca1}. To demonstrate that the attack described in \cite{OurP} does not work for the proposed scheme based on energy concealment, we take signal length of $N=31$  with an additional energy concealing variable, and try to decode the encrypted signal using the decoding strategy presented in  \cite{OurP}. From Fig. \ref{ca3} it is clear that the reconstructed signal does not leak any useful information for such a small length of an input signal. Therefore, theoretical and empirical evidence suggests that the proposed scheme is resistant against the ciphertext-only attack. 

 \begin{figure*}
	\centering
	\subfloat[]{\includegraphics [width=.13\textwidth] {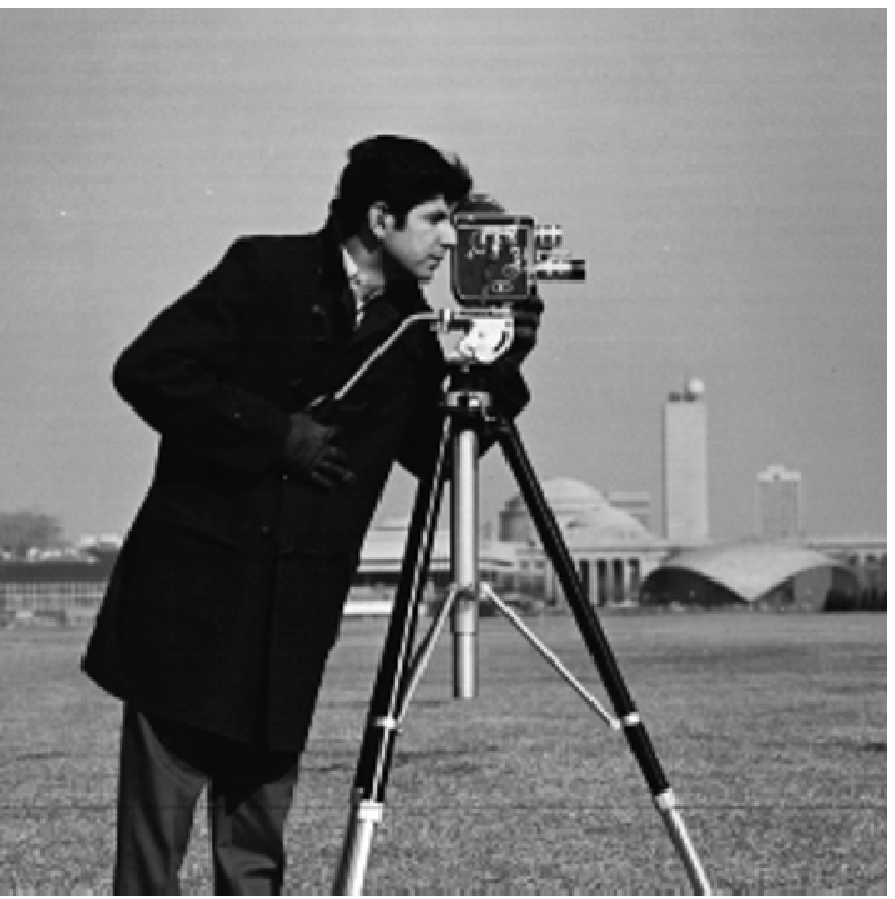}	\label{eos1}}
	\subfloat[]{\includegraphics [width=.13\textwidth]{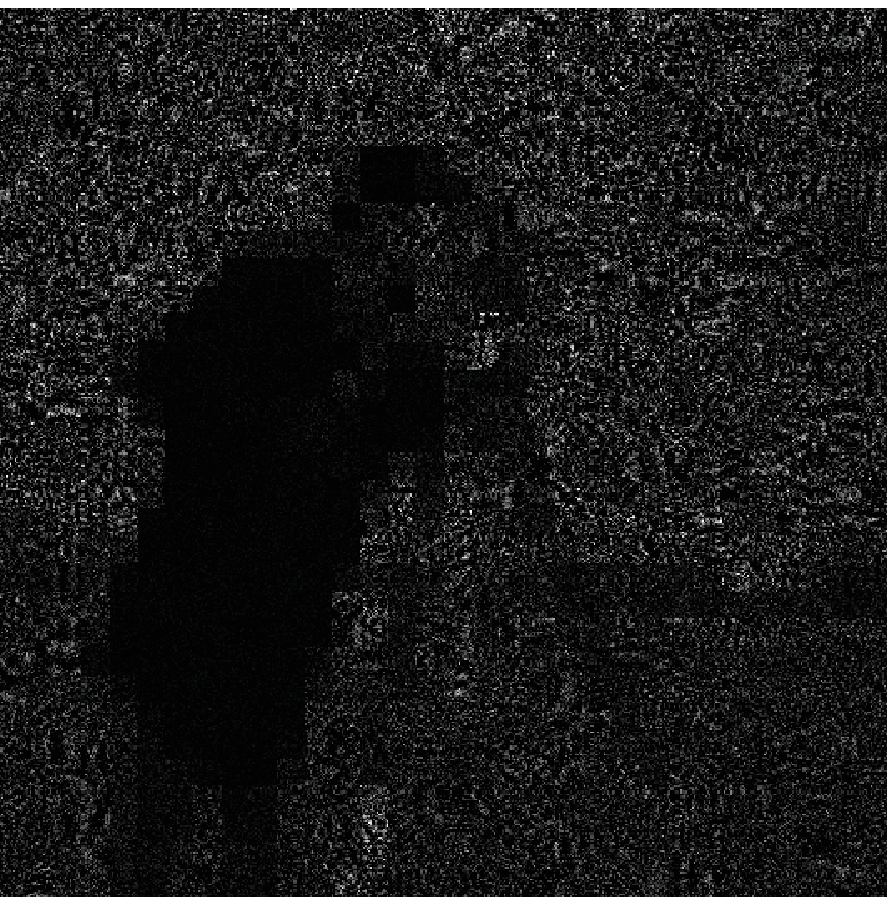}\label{eos2}}
	\subfloat[]{	\includegraphics [width=.13\textwidth]{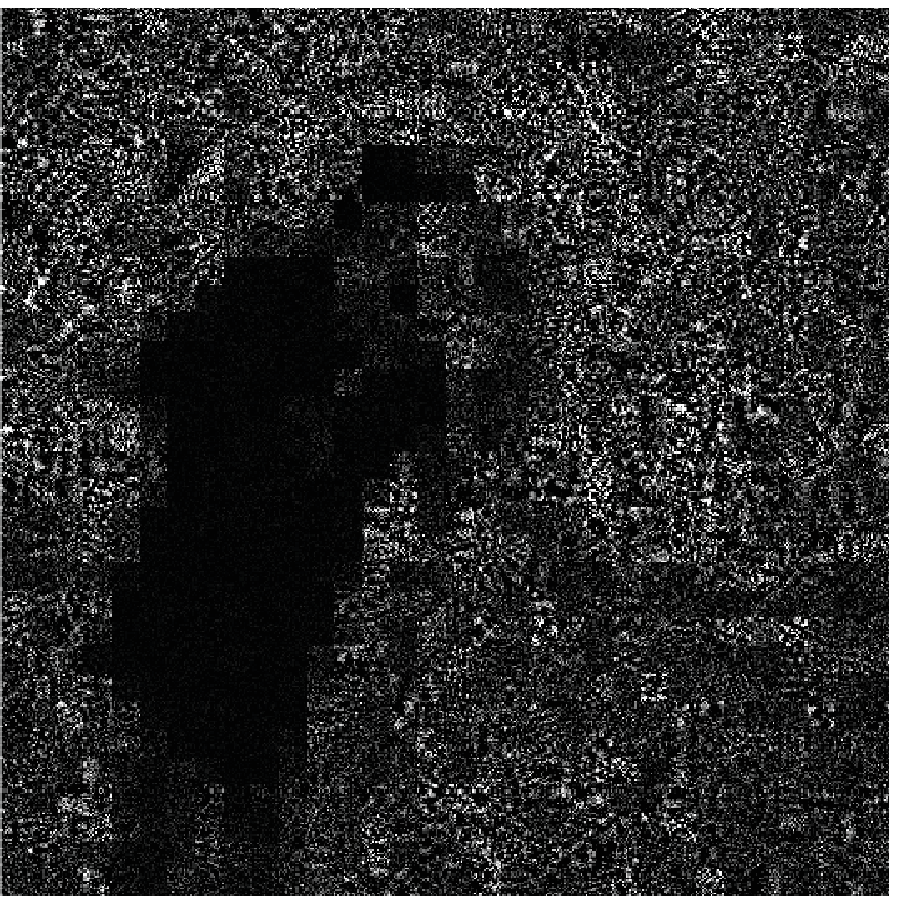}\label{eos3}}
	\subfloat[]{	\includegraphics [width=.13\textwidth]{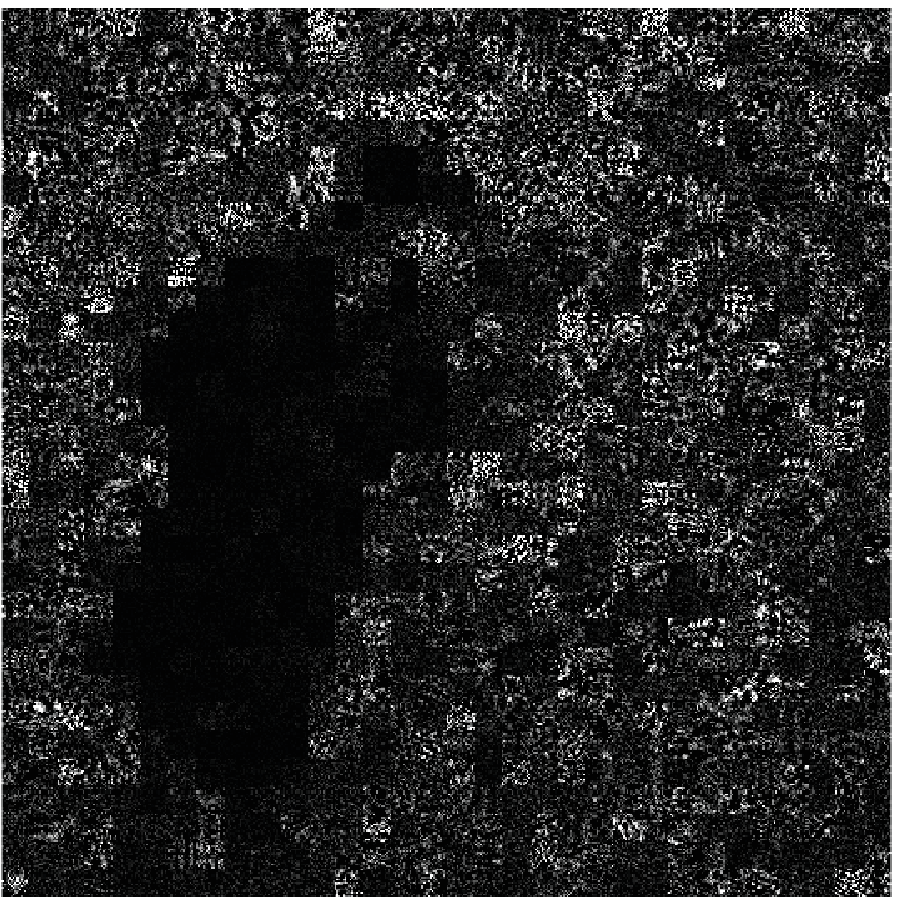}\label{eos5}}
	\subfloat[]{	\includegraphics [width=.13\textwidth]{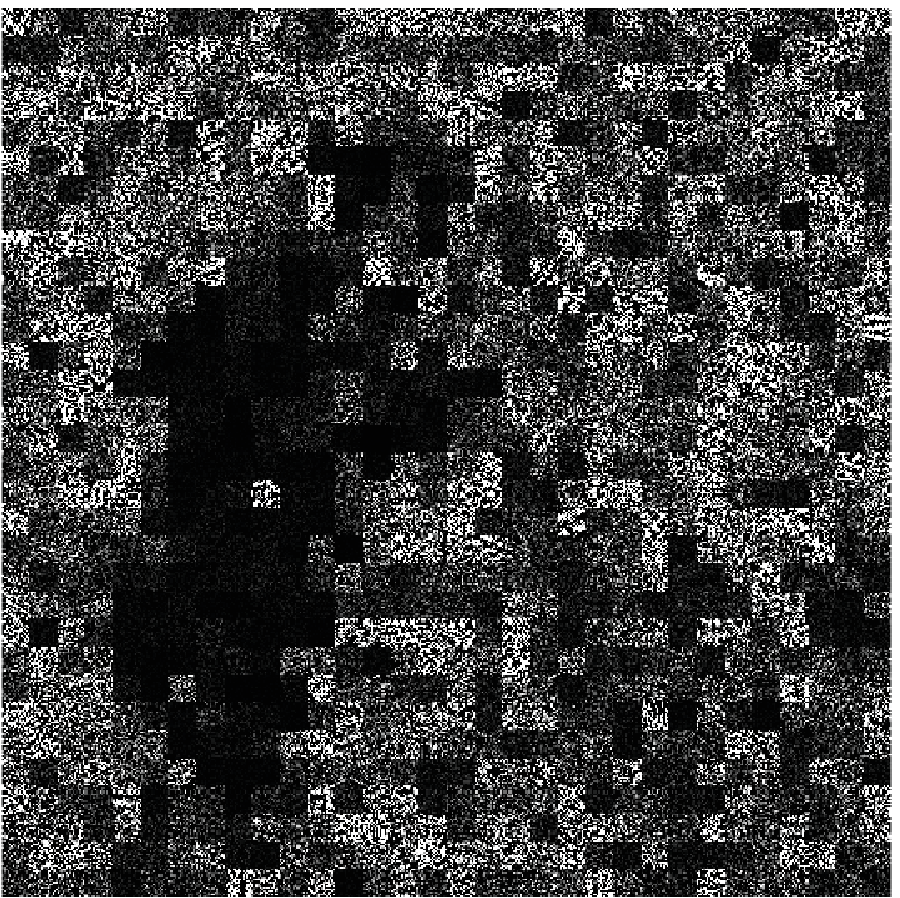}\label{eos6}}
	\subfloat[]{	\includegraphics [width=.13\textwidth]{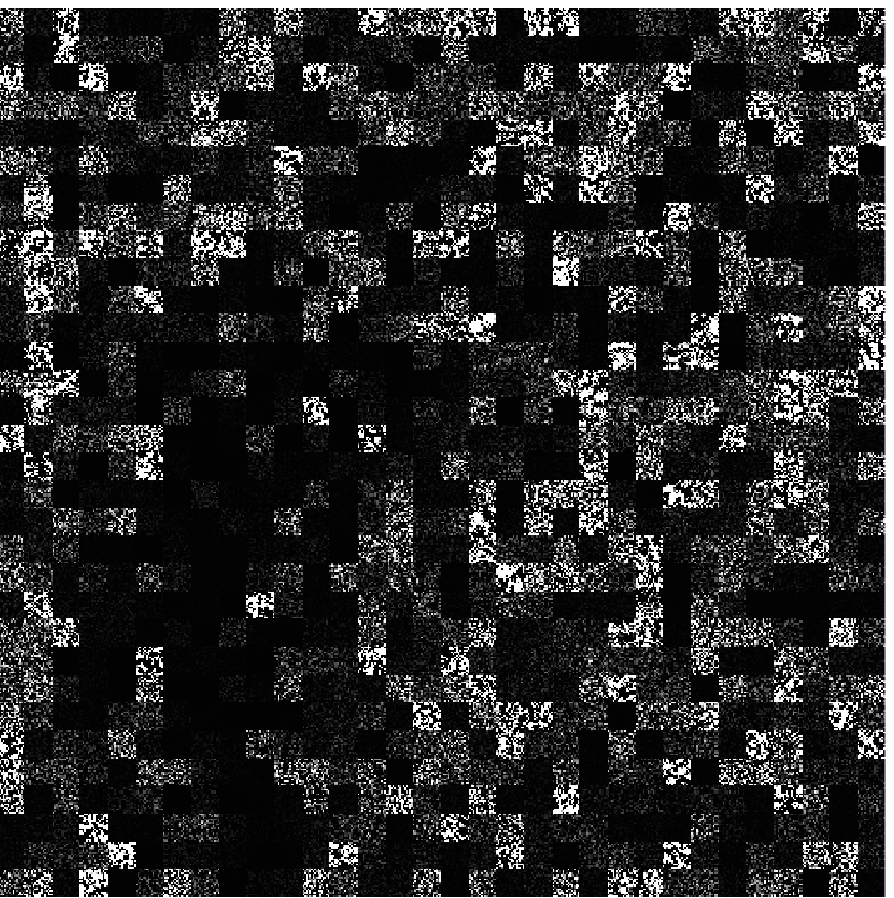}\label{eos7}}
	\subfloat[]{	\includegraphics [width=.13\textwidth]{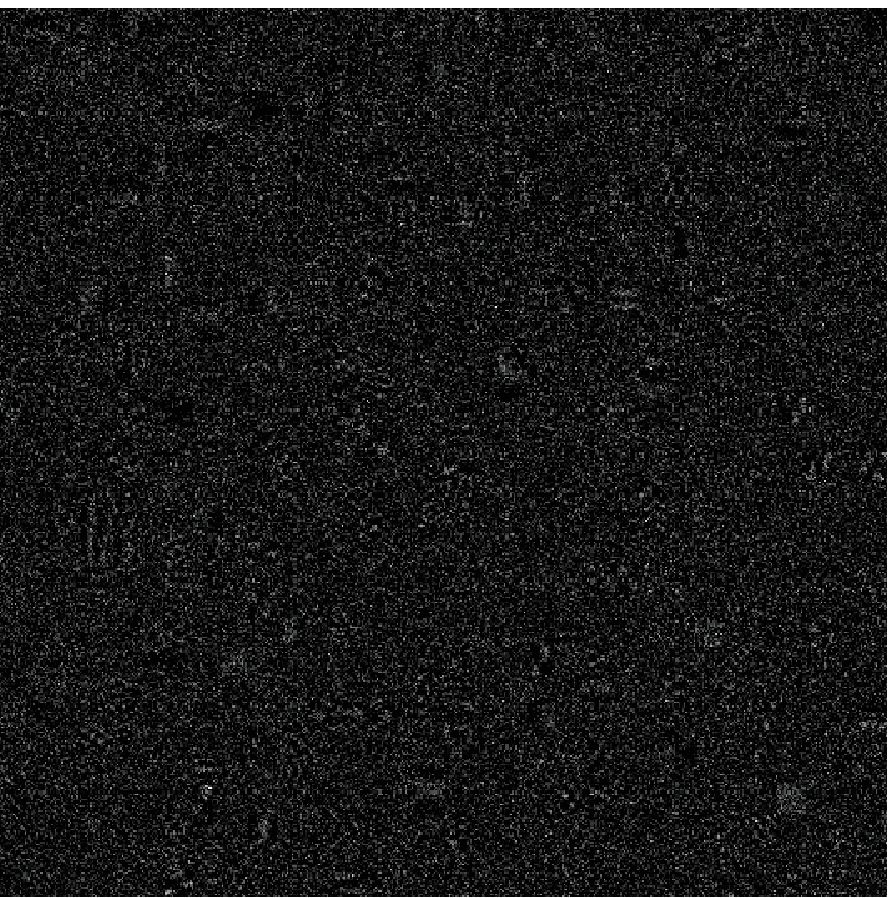}\label{eos8}}
	\caption{Application of ciphertext-only attack on computationally secure scheme, energy obfuscation scheme, and energy concealment encryption scheme.(a) Original image. (b) Reconstructed signal from OTS. (c) Reconstructed signal from EOS with $\sigma_a=.2$. (d) Reconstructed signal from EOS with $\sigma_a=.4$. (e) Reconstructed signal from EOS with $\sigma_a=1$. (f) Reconstructed signal from EOS with $\sigma_a=2$. (g) Reconstructed signal from energy concealment encryption scheme.}
\end{figure*} 
{In addition, we compare our scheme with the state-of-art energy obfuscation scheme (EOS)~\cite{EOS} under the ciphertext-only attack scheme from~\cite{OurP}. The authors of~\cite{EOS} have shown that the energy of the signal can be obfuscated by incorporating a random multiplier to the measurement as: $\mathbf{z}=a\mathbf{y}=a\mathbf{\Phi x}$, where $a\in\mathbb{R}$ is taken from a log-normal distribution $a\sim ln(0,\sigma^2_a)$. EOS is secure for large values of signal length and $\sigma^2_a$.} However, higher values of $\sigma_a$ will increase the magnitude of the measurement vector thereby increasing the signal dynamic range, hence may result in more storage for processing the measurement vector.

We implemented energy obfuscation algorithm in the MATLAB and performed its security analysis on image signals. For applying CS on an image, it is divided in  blocks of smaller size. Application of EOS on the $i^{th}$ block of the image is given as,
\begin{eqnarray}
\mathbf{z}_i=a_i\mathbf{\Phi}_i\mathbf{x}_i, \label{EOS}
\end{eqnarray}
where $\mathbf{x}_i$ is the vectorized $i^{th}$ block of the image and $\mathbf{\Phi}_i$ is iid Gaussian sensing matrix.    

 For security analysis of energy obfuscation scheme, we choose cameraman image of size $512\times512$ as shown in Fig.~\ref{eos1}. Encryption is performed on $16\times16$ image block using Eq. \ref{EOS} with $M=80$. We apply the attack method described in \cite{OurP} for discrete cosine basis and observe that energy obfuscation scheme is vulnerable to ciphertext-only attack. As it can be observed from Fig.~\ref{eos2}-\ref{eos7}, the adversary can recover plaintext information from ciphertext by applying ciphertext-only attack,  whereas the energy concealment encryption scheme does not leak any plaintext information as it can be observed from Fig.~\ref{eos8}.    
Apart from the possibility of ciphertext-only attack, energy obfuscation scheme requires extra resources to generate lognormal distribution at the IoT device and key management, which further introduces burden on communication protocol.  
 \subsection{Known/Chosen plaintext attack}
 In this subsection, the applicability of plaintext attack on the proposed scheme is discussed. The known plaintext attack  for binary sensing matrix is discussed in \cite{p1}. Authors claimed that for binary sensing matrix known plaintext attack is equivalent to the subset sum problem \cite{SSP} which is NP-hard. They also proved that it is computationally infeasible to guess the sensing matrix from the plaintext ciphertext pair because of many spurious solutions. Here, we use superincreasing sequences \cite{SIS} to show that a single pair of plaintext and ciphertext can reveal the complete sensing matrix when the sensing matrix entries are binary.     
 \begin{mydef}
 A	sequence of  numbers $\{b_1,b_2,\dots,b_N\} $ is superincreasing if $b_i \ge \sum_{j=1}^{i-1} b_j+1$ for $i>1$ and $b_1$ is any positive value.
  An example: $\{1,2,4,,8,...\}$ is a superincreasing sequence.
 \end{mydef} 
 \begin{myth}\label{th11} Given one pair of $\mathbf{y}$  and $\mathbf{x}$ such that $\mathbf{y}=\mathbf{\Phi}\mathbf{x}$. $\mathbf{\Phi}$ can be reconstructed correctly from $\mathbf{y}$, if $\mathbf{x}$ is superincreasing and $\mathbf{\Phi}$ is a Binomial sensing matrix. 
 \end{myth}
\textit{\hspace{3mm} Proof :} The $i^{th}$ row of sensing matrix is constructed by taking the $i^{th}$ element of $\mathbf{y}$. $y_i$ can be represented as,
\begin{eqnarray}
y_i=\sum_{j=1}^{N}\phi_{ij} x_j. \label{th1}
\end{eqnarray}
 $\mathbf{x}$ is superincreasing  for  $j>1$, and we have $x_j\ge \sum_{l=1}^{j-1} x_l+1$.

To estimate the value of $\phi_{ij}$ for $j=N$, it needs two steps.
\begin{itemize}
	\item  If $y_i>0$, this shows that $x_N$ is included in Eq. \ref{th1} so $\phi_{iN}=1$.
	Otherwise, $y_i<0$, this shows that $x_N$ is subtracted in Eq. \ref{th1} so $\phi_{iN}=-1$. 
	\item  To remove the effect of $x_N$, $y_i$ is updated in the following way,
	$y_i'=y_i-\phi_{iN}x_N$.
	
\end{itemize}

By assigning $y_i'$ to $y_i$ and repeating the step 1 and 2  for $j=N-1\text{ to }1$, the $i^{th}$ row of the sensing matrix is constructed. Therefore, by applying the above described procedure on each element of $\mathbf{y}$, $\mathbf{\Phi}$ can be  constructed correctly.

Theorem \ref{th11} shows that CS-based encryption schemes which use a weak random number generator to construct a Binomial sensing matrix are prone to chosen-plaintext attack because an adversary can deduce key from the sequence which constructs the Binomial sensing matrix \cite{l2, l3}. Our proposed scheme overcomes these weaknesses because we claim that retrieving the key from the  knowledge of sensing matrices is equivalent to solving a signal separation problem.

An adversary can break the proposed scheme, only if he first reconstructs the sequence which generates the sensing matrix. For an integer sensing matrix Theorem 1 does not apply directly but we assume that the adversary can learn the sequence generated by Eq. \ref{Gauss3} using a chosen-plaintext attack.  Then from this sequence the adversary reconstructs the LFSR sequences and NFSR sequence. An adversary can apply the Berlekamp
Massey algorithm \cite{BM} on the
learned LSFR sequences to get the key.  We claim that reconstructing LFSR sequences and NFSR sequence from  $\mathscr{F'}$ is a signal separation problem. We present the following theorem to show the difficulty of the signal separation problem. 
\begin{myth}\label{th2}
	Given a sequence $\mathscr{F'}$ of length $F$ generated by $L-1$ LFSRs and one NFSR using Eq. \ref{Gauss3}, the number of possible sequences to generate this sequence is given as:
	\begin{eqnarray}
	N_{Total}=\prod_{i=1}^{L-1} {L \choose i}^{r_{L-2i}}, \label{BF}
	\end{eqnarray}  
	where $r_{L-2i}$ is the number of times symbol $L-2i$ occurred in the sequence.
	Eq. \ref{BF} is lower bounded by 
	\begin{eqnarray}
         L^F\le N_{Total}. \label{BF1}
	\end{eqnarray}
\end{myth}
\textit{\hspace{3mm} Proof :} To generate symbol $L-2i$ we need   $L-i$ times $``1"$ and  $i$ times $``-1"$. Therefore the number of ways the symbol $L-2i$ is generated is given as  ${L \choose i}$. Symbol $L-2i$ is repeated for ${r_{L-2i}}$ times in the given sequence. Hence, the total number of possible ways of getting this particular symbol is  ${L \choose i}^{r_{L-2i}}$. The same logic can be applied for each symbol to get the total number of possible sequences which can generate the given sequence.

 We know that ${L \choose i} \ge L$ for $i=1$ to $L-1$. So from Eq. \ref{BF} we get 
\begin{eqnarray}
\prod_{i=1}^{L-1} {L }^{r_{L-2i}}\le N_{Total}. 
\end{eqnarray} 
Using $F=\sum_{i=1}^{L-1}r_{L-2i}$, we get the lower bound as in Eq. \ref{BF1}.

Theorem \ref{th2} gives us strong confidence that even if the sequence $\mathscr{F'}$   is given to an adversary. Reconstructing LFSR sequences is harder than brute-forcing the key. For example, each element of a sensing matrix for  $L=3$ takes value from the set $\{-1,1\}$. Each element of the sensing matrix can come from 3 different ways. e.g., symbol one can come from these three possibilities  $\{[1,1,-1],[1,-1,1],[-1,1,1]\}$. Hence, for a sensing matrix of size $10\times 32$ the possible number of LFSR sequences are $3^{320}$, which is more cumbersome than to apply brute force on each LFSR  and NFSR to get key.

\begin{myth}\label{th3}
	Given a sequence generator using Eq. \ref{Gauss1}, sequence  $\mathscr{F}$ is periodic with period $P$. The period of sequence  $\mathscr{F}$ can be  lower bounded by
	 \begin{eqnarray}
   lcm(P_1,\dots, P_{L-1})\le P,
	\end{eqnarray}
{when $max_i$(degree(LFSR$_i$))$> $degree(NFSR).}
\end{myth}
\textit{\hspace{3mm} Proof :} Sequence  $\mathscr{F}$ is generated by adding $L-1$ sequences generated from LFSRs and one sequence generated from nonlinear sequence generator. Therefore, sequence  $\mathscr{F}$ is periodic with  period $P$. From \cite{sm} we know that the period of a sequence generated by summing LFSRs is given as,
$
P'=lcm(P_1,\dots, P_i,\dots, P_{L-1})$,
where $P_i$ is the period of the $i^{th}$ LSFR.
The period of sequence  $\mathscr{F}$ is $P=lcm(P', P_{nl})$, where $P_{nl}$ is the period of nonlinear sequence. It is clear that $P'>P_{nl}$ which proves the claim.

The periodicity of the sensing matrix can be approximate using Theorem \ref{th3}. Assuming that sequence  $\mathscr{F}$ has period $P$ then the total number of extreme points in the full sequence is $\frac{P}{2^{L-1}}$. Therefore, the period of sequence $\mathscr{F'}$ is $P'=P-\frac{P}{2^{L-1}}$. 
For constructing a sensing matrix, $MN$ elements are required. Therefore, the repeatability of the sensing matrix is approximate $\frac{P'}{MN}$.
 For $L=11$ and LFSRs are having primitive polynomial degree in the order of $20$, the repeatability of sensing matrix is approximately $\frac{P'}{MN}\approx 2^{200}$ for $M=2^6$ and $N=2^8$. 

 \subsection{Reusability of sensing matrix}
  Due to the linearity in CS encryption, a scalar multiple of plaintext results in the corresponding multiplication in the ciphertext. For Binomial sensing matrix in some schemes \cite{powersmart} it is suggested that sensing matrix can be reused. However, from Theorem \ref{th11} it is clear that due to the presence of linearity in the encryption process, a sensing matrix cannot be used more than once  if the scheme has to be chosen-plaintext resistant. Note that in some IoT use cases, if  the chosen-plaintext attack is not feasible, then it is possible to reuse the sensing matrix. There is a tradeoff between security level and resource cost.

\section{Conclusion}
{In this paper, we prove that perfect secrecy is achievable for compressive sensing based one-time sensing using the Gaussian distributed sensing matrices. We design a cryptographic primitive using the combination of LFSR and NFSR to construct the approximately Gaussian distributed sensing matrix. The energy concealment encryption scheme is proven to be resistant against the cryptographic attacks.  We prove that retrieving the key from the sensing matrix construction sequence  is equivalent to solving a signal separation problem, which is harder than applying brute-force attack.  In addition, we show that the performance of the proposed scheme is equivalent to the one time sensing scheme.}

\section*{Acknowledgement}
This work is supported by Innovation Fund Denmark (Grant no.8057-00059B.)  and DIGIT center Aarhus University.

\end{document}